\documentclass[
               ,final
              ]{aipproc}

\layoutstyle{6x9}

\usepackage{epsfig}  %% To include eps figures 
\usepackage{comment} %% For large regions of commented text
\usepackage{amssymb} %% For the AMS symbols
\usepackage{amsmath}     %% Utilizar herramientas matematicas extra
\usepackage{natbib}

\begin{document}  

\title{Magnetic Edge States in Graphene}
       

\author{Gabriela Murgu\'ia}{
%\email{murguia@ciencias.unam.mx}
  address = {Departamento de F\'isica, Facultad de Ciencias, 
  Universidad Nacional Aut\'onoma de M\'exico, \\
Apartado Postal 21-092,  04021, Distrito Federal, M\'exico.}
}

%%% Abstract %%%   
\begin{abstract}  
Magnetic confinement in graphene has been of recent and growing interest because its potential applications in nanotechnology. In particular, the observation of the so called magnetic edge states in graphene has opened the possibility to deepen into the generation of spin currents and its applications in spintronics. We study the magnetic edge states of quasi-particles arising in graphene monolayers due to an inhomogeneous magnetic field of a magnetic barrier in the formalism of the two-dimensional massless Dirac equation.
We also show how the solutions of such states in each of both triangular sublattices of the graphene are related through a supersymmetric transformation in the quantum mechanical sense.
\end{abstract}  
%%% End Abstract %%%   

\maketitle

%%% Section: Introduction %%%   
\section{Introduction}   

 Graphene is a novel two dimensional material that has opened a new bridge of common interests between the condensed matter and high energy physics communities. It consists of a single atomic monolayer of carbon atoms arranged in a honeycomb hexagonal crystal lattice that can be described through two triangular Bravias sublattices $A$ and $B$.
Some of its properties of interest for the nanoelectronics are its high electrical and thermal conductivity as well as its high elasticity and toughness, making it a lightweight and flexible material that could be used instead of silicon in some semiconductor electronic devices.
The structure of the energy bands of graphene is such that the dispersion relation is a linear one, implying that the charge carriers have a zero effective mass. Then, the graphene can be studied through the Dirac equation in $(2 +1)$D in the zero mass limit, where the two irreducible representations of the Dirac matrices describe two species of particles, one for each triangular sublattice and labeled by a pseudo-spin originated by their interactions with the lattice~\cite{Semenoff-Vicenzo}. 
Interacting with an external magnetic field, the electrons in the crystal lattice of graphene are subjected to an effective potential that has a minimum below the zero of the external potential~\cite{DeMartino}. This implies the existence of bound states solutions, giving rise to some phenomena of confinement like the so called magnetic edge states, magnetic quantum dots and magnetic spin currents, all of them of growing interest in  the nanotechnology.
Here we obtain the magnetic edge states in graphene for a static external magnetic barrier making use of the supersymmetric character, in the quantum mechanical sense (SUSY-QM), of the Dirac equation for external and time-independent magnetic fields~\cite{Cooper}.

%%% Section: Dirac Equation in (2+1)D and SUSY-QM %%%  
\section{Dirac Equation in (2+1)D and SUSY-QM}  

In the context of the relativistic quantum mechanics,  a supersymmetric Hamiltonian
% $H_\text{SUSY}$ 
can be written in the general form as
$H_\text{SUSY} = \{Q,Q^\dagger\},$
with the supercharges $Q$ and $Q^\dagger$ being nilpotent operators.
% which comute with $H_\text{SUSY}$.
%%
We are interested in the $(2+1)$D Dirac equation in the massless limit,
\begin{equation}
(\gamma \cdot \Pi) \Psi({\bf x},t) = 0,
\label{eq:EcDirac}
\end{equation}
with $\Pi_\mu = i\partial_\mu + eA_\mu$\footnote{We work in natural units such that $\hbar = \tilde{c} = 1$, being $\tilde{c}$ the Fermi velocity, which for graphene plays the role of the speed of light and is two orders of magnitude smaller than $c$ ($\tilde{c} \sim c/300$)~\cite{Geim-Peres}.} and $\Psi({\bf x},t) = \psi({\bf x}) e^{-iEt}$.
In this case, there are two irreducible representations for the $\gamma$ matrices given in terms of the $2\times2$ Pauli matrices. We will work with the Jackiw representation, for which the first irreducible representation, labeled as $A$, corresponds to the set
$\{\gamma^{0} = \sigma_3, \gamma^{1}  = i\sigma_1, \gamma^{2}  = i\sigma_2\},$
and the second irreducible representation, $B$, is given by
$\{\gamma^{0} = \sigma_3, \gamma^{1}  = i\sigma_1, \gamma^{2}  = -i\sigma_2\}.$
 
Taking the square of the $(\gamma \cdot \Pi)$ operator in Eq.~(\ref{eq:EcDirac}), $(\gamma \cdot \Pi)^2 \Psi$, a time-independent Schrödinger-type equation for $\psi({\bf x})$ arises in terms of a SUSY-QM Hamiltonian, $H_{\sigma, R}= \{Q_R,Q_{R}^{\dagger}\} = -(\gamma \cdot \Pi)^2 + \Pi_{0}^{2}$, which depends on both, the spin of the particles ($\sigma$) and on the irreducible representation ($R=\{A,B\}$) of the Dirac matrices, as
 $H_{\sigma, R} \psi({\bf x}) = E^{2} \psi({\bf x}).$
The supercharges, in each irreducible representation, are 
$Q_{R} = (\Pi_1 + \Pi_2) \sigma^{+}_{R}$
and  
$Q^{\dagger}_{R} = (\Pi_1 - \Pi_2) \sigma^{-}_{R}$,
with $\sigma_{A}^{\pm} = (\sigma^1 \pm i \sigma^2)/2 = \sigma_{B}^{\mp}$.
There are two corresponding effective potentials $V_{\sigma,R}$, one per each irreducible representation, called SUSY partner potentials.
For an external static magnetic field described in the Landau gauge by the tri-potential $A^\mu = (0,0,W(x))$, these are given in terms of the superpotential $\bar{W} = e W(x) + p_2$ as $V_{\sigma,R} = \bar{W} + (-1)^R \sigma \bar{W}^\prime$ with $\bar{W}^\prime = dW(x)/dx$, $p_2$ being the component of the momentum along the $y$ direction and $R=1,2$ for the irreducible representation $A$ and $B$ respectively. 
 
As a consequence of the SUSY-QM character of $H_{\sigma, R}$, the Dirac equation~(\ref{eq:EcDirac}) for any external static magnetic field, 
reduces to a Pauli one with effective mass $m=1/2$ and gyromagnetic ratio $g=2$~\cite{Matias_1992}.
Due to the fact that there are two irreducible representations for the Dirac matrices in the  $(2+1)$D case, there is also a direct relation between the solutions of the wave function for different spin eigenvalues, which shows itself through the $\sigma^{\pm}_{R}$ operators.

%%% Section:  %%%  
\section{Magnetic Barrier}

Applying the formalism of SUSY-QM for a magnetic barrier of length $d$ perpendicular to the plane of a graphene monolayer, given by $B(x) = B_0 \Theta(d^2-x^2)$, each one of the effective potentials $V_{\sigma,R}$ corresponds to a truncated harmonic oscillator that in general is not symmetric. The condition for the existence of bound states solutions, referred in the literature as {\it edge states} (because they are localized at the borders of the external applied potential),  in this case is given if the energy $E$ of the charge carries satisfies the relation ${\ell}^{2}_{B} E^2 < V_{0}^{(-s)}$, being $s = \text{sign}(p_2 {\ell}_{B})$, $V_{0}^{(\pm)} = (p_2 {\ell}_{B} \pm d/{\ell}_{B})$  and ${\ell}^{2}_{B} = \sqrt{\hbar c/(e B)}$ is the magnetic length. 

\begin{center}
\begin{figure}[ht]
\includegraphics[width=0.45\linewidth]{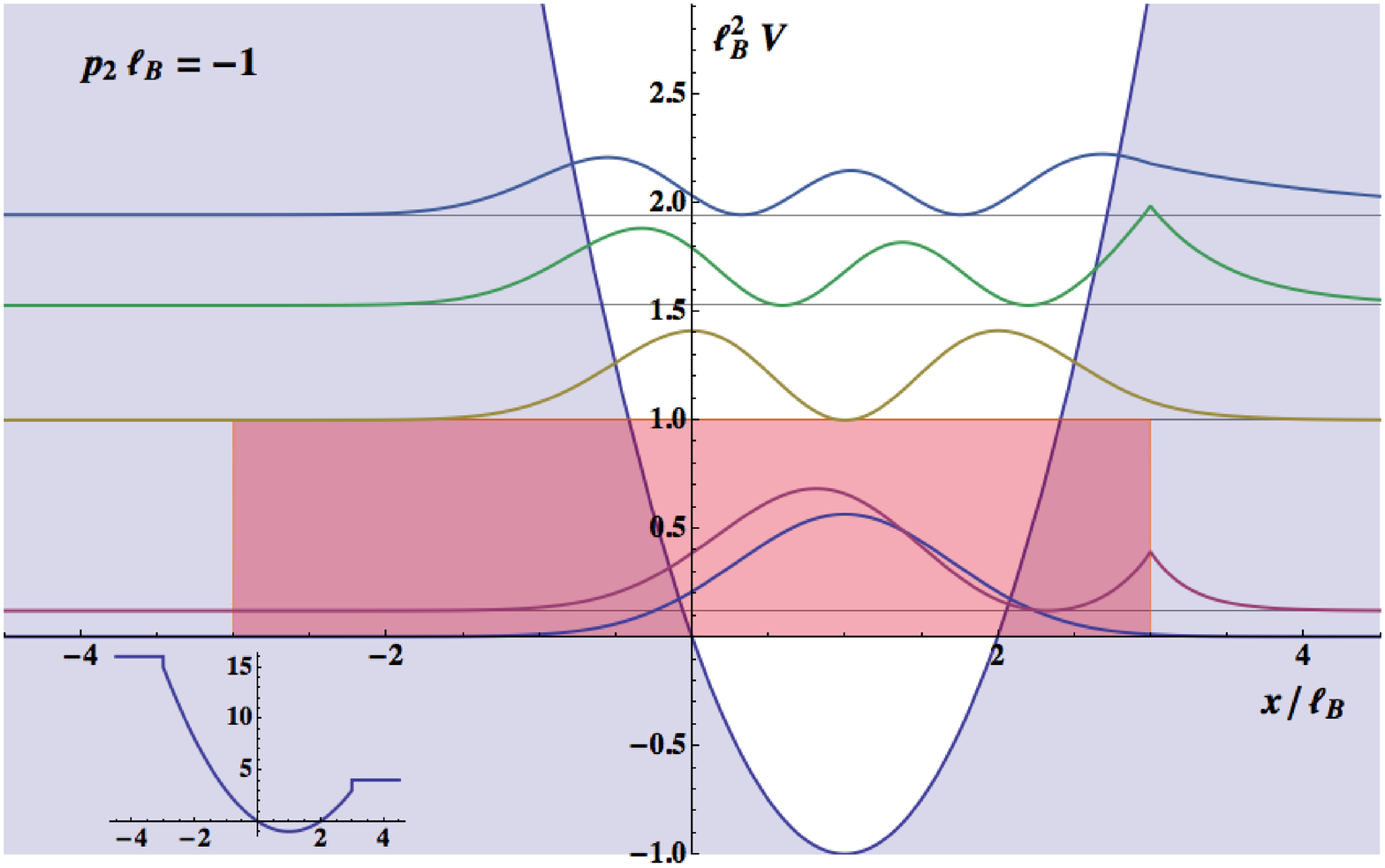}
\hspace{.1\linewidth}
\includegraphics[width=0.45\linewidth]{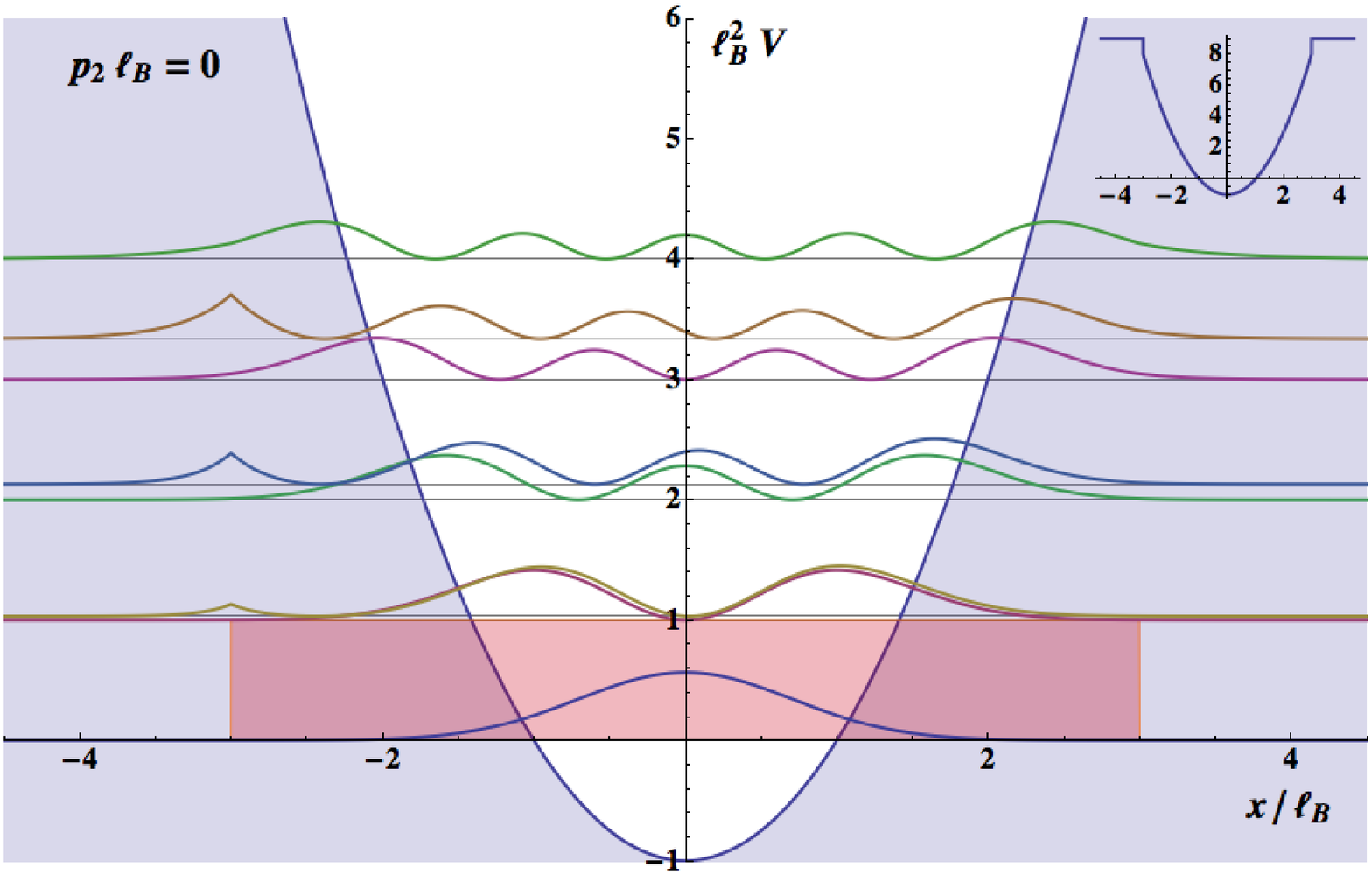}
 \caption{The probability density function $|\psi(x/\ell_B)|^2$ of the magnetic edge states for a magnetic barrier of length $d=3\ell_B$ for $(-1)^{R}\sigma = -1$ and $p_2 \ell_B=$-1.0 ({\it left}), 0.0 ({\it right}). The magnetic barrier, the effective truncated harmonic oscillator, as well as a {\it zoom} of this last in the energy region of the bound states, are schematically shown in both cases.}
 \label{figure:plots}
\end{figure}
\end{center}

Figure~\ref{figure:plots} shows the probability density function $|\psi(x/\ell_B)|^2$ for the calculated magnetic edge states for a magnetic barrier of length $d = 3 {\ell}_{B}$ for two values of the ${\ell}_{B} p_2$ parameter. The spin direction has been selected as $(-1)^R \sigma = -1$.

%%%%%%%%%%%%%%%
% figure::plots
%%%%%%%%%%%%%%%

%%% Section: Conclusions %%%
\section{Concluding Remarks}  

SUSY-QM applied to graphene in a static and inhomogeneous external magnetic field, allows to map the wave functions of the charge carriers in one of the two triangular sublattices, with definite induced pseudo-spin, into the other. In this way the solutions of the $(2+1)$D Dirac equation in each of the two irreducible representations are closely related.
In the zero mass limit, the relativistic description of the system has been reduced to a non-relativistic one characterized through a time-independent Schrödinger-type equation.
Under this formalism, the magnetic edge states of graphene for a magnetic barrier were obtained.

%%% References %%%

\end{document}